\documentstyle[twocolumn,aps,prl]{revtex}

\def\simge{\mathrel{%
   \rlap{\raise 0.511ex \hbox{$>$}}{\lower 0.511ex \hbox{$\sim$}}}}

\draft

\include{psfig}
\begin{document}

\twocolumn[\hsize\textwidth\columnwidth\hsize\csname
@twocolumnfalse\endcsname

\title{Cross-Correlation of the Cosmic Microwave Background with
Radio Sources: Constraints on an Accelerating Universe}
\author{S.P. Boughn}
\address{Department of Astronomy, Haverford College, Haverford, PA 19041
{\tt sboughn@haverford.edu} }
\author{R.G. Crittenden}
\address{DAMTP, University of Cambridge, Cambridge CB3 9EW, UK
{\tt r.g.crittenden@damtp.cam.ac.uk}}

\maketitle 
\begin{abstract}
We present a new limit on the cosmological constant based on the absence
of correlations between the cosmic microwave background (CMB) and
the distribution of distant radio sources.  In the cosmological constant-cold 
dark matter ($\Lambda$CDM) models currently favored, such
correlations should have been produced via the integrated Sachs-Wolfe
effect, assuming that radio sources trace the local ($z\sim1$) matter
density.  
We find no evidence of correlations between the COBE 53Hz microwave map
and the NVSS 1.4 GHz radio survey, and obtain an upper limit for the
normalized cross-correlation of  $\langle \delta N \delta T \rangle /
\sqrt{\langle \delta N^2 \rangle \langle \delta T^2 \rangle} \leq 0.067$
at the 95\% CL.  This corresponds to an upper limit on the cosmological 
constant of $\Omega_\Lambda \le 0.74,$ which is in marginal agreement with 
the values suggested by recent measurements of the CMB anisotropy and
type-IA supernovae observations, $\Omega_\Lambda \simeq 0.6-0.7.$ 
If the cosmological model does lie in this range, then the integrated
Sachs-Wolfe effect should be detectable with upcoming CMB maps and
radio surveys.
\end{abstract} 
]


Recent observations of supernovae light curves \cite{SN} suggest that the
expansion of the universe is accelerating rather than decelerating.
Combined with evidence from the cosmic microwave background (CMB)
\cite{boom,max,dasi}   
and a number of other observations \cite{bahcall}, this suggests the universe  
is spatially flat and dominated by a cosmological constant,
$\Omega_{\Lambda} \equiv \Lambda/3H_0^2 \simeq 0.6-0.7$, where $H_0$ is the
Hubble expansion parameter.  Such a low value of $\Lambda$ is difficult to
explain from fundamental grounds, so it is vital that we try to confirm
this result by other means.


CMB anisotropies can arise via the 
integrated Sachs-Wolfe effect (ISW)\cite{sw} as the photons travel through the
time-dependent gravitational potentials of collapsing structures.  
One consequence of a large cosmological constant is that such 
time-dependent potentials exist even
on very large scales where the collapse is linear, which is not the case
for a flat, matter dominated universe. 
These fluctuations are likely to be small
compared to those imprinted at the surface of last scattering (redshifts  
$z \sim 1000$) and are difficult to detect directly,  however, they can
be observed by looking for spatial correlations between the CMB and the   
nearby matter density\cite{ct,kam}.  
This requires a probe of the matter density out
to redshifts of $z \sim 2$ and suggested candidates include radio
galaxies, quasars and the x-ray background.

Other processes can also lead to correlations between the CMB and the 
local matter density. 
These include gravitational lensing, scattering from hot electrons
(the Sunyaev-Zeldovich effect) and photons passing through the time-dependent
potentials of non-linear collapsing structures (the Rees-Sciama effect).
While the study of these effects can also benefit   
from cross correlation analyses \cite{ps}, the ISW effect is unique in that it
occurs on very large scales ($\theta > 1 ^{\circ}$) where the fluctuations
are simple and linear.

In the first attempt to detect this  
effect, Boughn, Crittenden, and Turok \cite{bct} cross-correlated the 
CMB with the hard ($>$2 keV) x-ray background, which is thought to 
arise primarily from active galactic nuclei out to a redshift of 
$z \sim 4$\cite{com}. 
A COBE-DMR CMB map \cite{COBE} was cross-correlated with the HEAO1
x-ray map \cite{bol} and no significant correlation between 
the two maps was found. 
The interpretation of this result is difficult since there has been no 
unambiguous measurement of correlations in the x-ray background, 
so the x-ray bias factor, the extent to which x-rays trace matter,
is largely unknown.  For a large x-ray bias, $b_x \sim 4$, the
implication is that  
$\Omega_\Lambda \leq 0.5$, in conflict with the currently favored 
$\Lambda $ cold dark matter ($\Lambda$CDM) cosmological model;
for no biasing, $b_x = 1$, the limit is much weaker: 
$\Omega_\Lambda \leq 0.95$ \cite{bct}. 
The true bias may well be time dependent and will not 
be known until the structure in the x-ray background is definitively detected. 

  
Here we attempt to detect the  
ISW effect by cross-correlating the CMB with a deep radio source
survey.  While the Poisson noise due to the finite number density of
radio sources is relatively large, this analysis has the advantage that
the clustering properties of the radio sources have been measured and
so the bias factor for these sources can be deduced within the 
context of a particular cosmological model.


The small correlations introduced by the ISW effect 
can be contaminated by accidental correlations of the locations
of radio sources with the relatively large CMB fluctuations
that originate at higher redshift.
To minimize these accidental correlations,  
it is necessary to average over many statistically independent 
regions of the sky.  
Since the ISW correlations are on 
angular scales of several degrees, 
this means using surveys that cover as much of the sky as possible,  even if 
they have more noise than smaller area surveys. 
Thus, in the radio we use the NVSS radio source 
survey (82\% sky coverage) and for the CMB we use the 4-year COBE 53GHz map
(full sky coverage.)
The 
31 GHz and 90 GHz companion maps both have significantly higher noise,
as do the combination maps constructed by the COBE team to minimize Galactic
emission.  
Above a Galactic latitude of $| b_{gal}|>10^\circ$,
the NVSS source counts are essentially all 
extra-Galactic.
Prior to cross-correlating the two data sets, a secant law Galaxy model
and a dipole (the largest structure in the 53 GHz map) were fit and
removed.  The effect of these corrections was minimal. 

The NRAO VLA Sky Survey (NVSS) is a $2.5\times 10^{-3}$ Jansky ($2.5 mJy$) 
flux limited survey
at a frequency of 1.4 GHz \cite{nvss}. 
It is complete for declinations 
$\delta > -40^{\circ}$ and contains
$> 1.8\times  10^6$ sources, with a 
mean source number density of 51.8 per square degree.  
While the distances of individual sources are largely unknown, typical
luminosity function models (e.g.,\cite{condon,dp}) indicate they are 
distributed in the redshift range $0 \leq z \leq 2$ with a
peak in the distribution at $z \sim 0.8$.  This distribution, combined 
with the nearly full-sky coverage, renders the NVSS survey an ideal 
matter density probe with which to investigate the ISW effect.  

The NVSS survey was converted into a map 
using an equatorial quadrilaterized spherical cube projection \cite{quad},  
the standard format for the COBE data.  
Those $1.3^\circ \times 1.3^\circ$ pixels  
that were only partially contained in the survey 
region were omitted.  
In making these maps, we discovered that 
the surface density of sources varied by $\sim \pm 5\%$
within several declinations bands.  The more prominent of these 
bands coincide with discontinuous changes in the rms noise levels in 
the NVSS survey \cite{nvss}.  To correct for this, random sources were 
added to or subtracted from each pixel to eliminate the band structure. 
The resulting 
map shows no declination dependent structure at level 
of $< 1\%$.  For comparison, the Poisson noise per pixel is $\sim 11\%$.

In order to exclude Galactic sources as well as nearby 
clusters of galaxies, the region within
$\pm 10^{\circ}$ of the Galactic plane was removed as were
29 regions that contained pixels with source counts greater
than $4\sigma$ ($43\%$) above the average.  This procedure cleaned
the map of 10 ``objects'' located more than $10^{\circ}$
from the Galactic plane.  Among these
are the Orion Nebula and the nearby Virgo, Perseus, and Fornax 
clusters.  That the moderately nearby, rich Coma cluster of galaxies 
was not one of the regions cut indicates to us that this windowing
removed only Galactic and nearby extra-galactic sources. 

After the cleaning and correction operations, 
the map was repixelized by combining groups of four pixels into larger 
$2.6^{\circ} \times 2.6^{\circ}$ pixels.  The angular resolution of
the COBE map is $\sim 7^{\circ}$ and this coarser pixelization is
the standard format provided for the COBE DMR data sets.
Each of the larger pixels was
assigned a weight ${\rm w_i}$ corresponding to the number of subpixels
contributing to it and its value was multiplied by $4/{\rm w_i}$ to
achieve the proper normalization.  The radio source auto-correlation
function (ACF) was computed according to
\begin{equation}
 \omega_{RR}(\theta) = \sum {\rm w_i}(N_i-\bar{N}) {\rm w_j} (N_j-\bar{N}) 
\left/ 
\sum {\rm w_i w_j}, \right.
\end{equation}
where the sums are over all pairs of pixels with angular separation
$\theta$, $\bar{N}$ is the mean number of sources 
and ${\rm w_i}$ is the pixel weighting factor.   The measured ACF is 
displayed in Figure 1 with error bars that were deduced from Monte Carlo 
simulations. 

\begin{figure}[htb]
\vspace{-0.4in} 
\centerline{\psfig{file=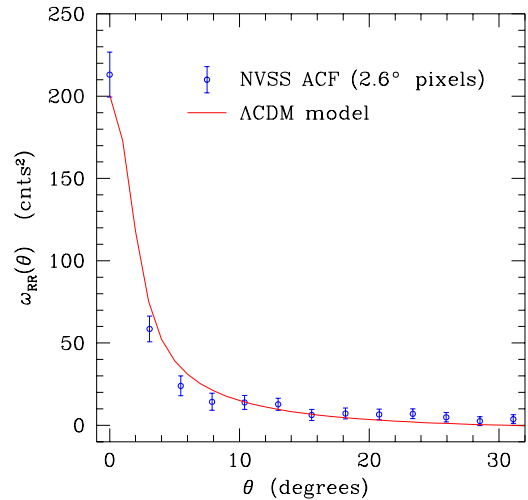,height=3.5in,width=3.5in}} 
\vspace{-0.3in} 
\caption{The auto-correlation function of the NVSS and the predictions from 
a $\Lambda$CDM theory, where $\Omega_\Lambda = 0.7$ and $h=0.7$.  
The inferred linear bias factor is $b_R=1.6$.  The theory has been 
convolved with the pixel window function.  
Note that the errors are highly correlated. 
}
\label{fig-acf}
\end{figure}

The effect of the above mentioned declination band corrections on 
the ACF was considerable.  For $\theta > 0$ the ACF of the 
corrected map is $\sim$ 2--3  times smaller than that of the
uncorrected map.  This is not unexpected given the prominent band 
structure in the uncorrected map.  To test the sensitivity of the ACF 
to our corrections, more aggressive band removal was effected by
adding (and subtracting) sources at random from even more narrow
declinations bands ($8^\circ$, $4^\circ$, and $2^\circ$) in order 
to force the mean number densities in these bands to be equal. 
In all such cases, the ACF's were not significantly different 
from that of Figure 1.  We conclude that our ACF is reasonably robust.

One can determine the radio source bias factor by comparing the measured 
ACF to the mass density correlation predicted theoretically   
assuming the COBE normalization and a given cosmological model. 
The curve in Figure 1 
is the prediction of a sample $\Lambda$CDM cosmology  
convolved with window functions 
for the quad-cubed pixelization. 
The inferred radio bias parameter depends on the exact cosmology, as is shown 
in Figure 3, but we find typical values in the range $b_R = 1.3-1.6.$  
Since we compare on very large angular scales where the fluctuations are linear,
the inferred biases are independent of the vagaries of 
non-linear structure formation.  

An essential ingredient in interpreting the auto- and 
cross-correlations is the source number-redshift distribution, $dN/dz$. 
The distribution used in Figure 1 was derived from 
a radio source luminosity function (LF)  
of Dunlop and Peacock (mean-$z$ model 1) \cite{dp}.  
This LF is
consistent with flux limited number counts of 
the present NVSS survey as well as with recent, deep
redshift surveys \cite{maglio}.  The mean-$z$ model 3 of Dunlop 
and Peacock and the model of Condon \cite{condon} 
are also reasonably consistent with the data and were also
considered.  
The bias factors deduced from these latter
two LFs are within $15\%$ of that deduced above, which indicate  
that the results are not overly sensitive to the details of 
$dN/dz$ or the LF. 

We can compare our results to the ACF of the FIRST radio survey,
which was determined by Cress et al. \cite{cress}  and Magliocchetti et al. 
\cite{maglio2}  
to be $\omega_{RR}(\theta) 
\simeq 1-2 \times 10^{-3}/\theta$ for $0.01^\circ \leq 
\theta \leq 4^\circ$ where $\theta$ is in degrees.  While the FIRST survey is
somewhat deeper ($\sim 1 mJy$) than the NVSS survey, the luminosity function
is such that the redshift distribution of these sources is nearly identical to 
that of the NVSS sources \cite{nvss}.  Therefore, one expects the 
ACFs for these two surveys to be nearly the same.  When converted to the
the pixelization used in this
paper, these results are consistent with that in Figure~1.


The cross-correlation function (CCF) of the radio and CMB maps 
was computed using
\begin{equation}
\omega_{RT}(\theta) = \sum {\rm  w_i} (N_i-\bar{N}) (T_j-\bar{T})
/\sigma_{j}^2\left/ 
\sum {\rm w_i/\sigma_{j}^2} \right. 
\end{equation}
where $T_j$ and $\sigma_{j}^2$ are the
temperature and the noise variance of the $j^{th}$ pixel in the CMB map,  
$\bar{T}$ is the mean CMB temperature, 
and the radio variables are defined as before. 
The COBE instrument noise 
is significant and the pixels were weighted accordingly
in Eq. 2.  However, if the pixels are weighted uniformly,
the resulting CCF does not change significantly.

While there may be unaccounted for systematics, 
it seems unlikely that these will be correlated 
in two such disparate maps.  This is born out by
the insignificant differences in the cross-correlation functions 
computed both with and without correcting the NVSS map for the 
declination band structure and with and without correcting the CMB
map for both a secant law Galaxy and a quadrupole moment.  

The CCF is plotted in Figure 2 and is consistent with the data 
sets being uncorrelated.  The effect of removing the dipole 
and secant law Galaxy model from the CMB map was investigated 
by cross-correlating a set of Monte Carlo simulations in which 
a small correlated component having the same profile as the 
$\Lambda$CDM CCF was added to larger components that match the 
ACFs of the two maps.  The results indicate that the dipole/secant 
law corrections significantly attenuate the CCF but only for 
separations, $\theta \ge 15^{\circ}$.  Therefore, these
data were ignored in the subsequent analysis. 
The error bars in Figure 2 were computed from the Monte Carlo
trials and are highly correlated.

\begin{figure}[htb]
\vspace{-0.4in} 
\centerline{
\psfig{file=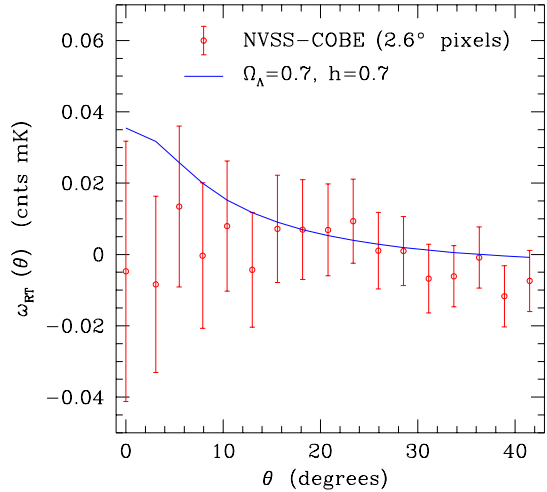,height=3.5in,width=3.5in}}
\vspace{-0.3in} 
\caption{The cross-correlation of the NVSS 
with the COBE microwave background 53GHz map.
The errors are highly correlated.}
\label{fig-ccf}
\end{figure}

The theoretical predictions for the cross correlations are 
calculated following the formalism of Refs. \cite{ct,bct}. 
In order to simplify comparison to the data, we consider a single 
family of possible correlation functions having the shape of that in Figure 2
but with variable amplitude. 
This family closely matches  
the correlations predicted for the range of $\Lambda$CDM 
cosmological models we consider. 
The 95\% CL (1.65 $\sigma$) upper limit for the zero lag correlation is
$w_{RT}(0) < 0.038$ while 
$w_{RT}(0) < 0.048$ at 98\% CL (2.05 $\sigma$.)   

The cosmological implications of this bound are summarized in Figure 3, where 
we plot the allowed region in the $H_0-\Omega_m$ plane, where  
$\Omega_m$ is 
the fraction of the critical density in both baryons and dark matter. 
Also shown on the plot are contours of constant radio bias and contours of
constant $\Gamma \equiv \Omega_m h$
($h \equiv H_0/100 km s^{-1} Mpc^{-1}$),  
which determines the shape of the matter power spectrum. 
To first order, the predicted cross-correlation depends only on the 
value of the cosmological constant, and models with $\Omega_\Lambda 
\simge 0.7$ are excluded.  However, there is some dependence on the Hubble 
parameter because the radio auto-correlation includes contributions 
from smaller scales than generally contribute to the cross-correlation. 
Thus, models with more small scale power (large $\Gamma$, $H_0$) tend 
to have smaller biases and predict less cross-correlation than models 
with lower $\Gamma$ and $H_0$.  

Our results have thus far assumed a constant linear bias factor, but in 
principle bias could be time and even scale dependent.  Simple models 
of linear bias evolution \cite{TDB} indicate that it is 
tied to the evolution of the linear growth factor $D(t)$ 
(normalized to be unity today):  $b_0 - 1 = (b(t) - 1)D(t).$ 
Thus, models that are presently positively biased 
were even more biased in the past.   
We do not expect this time evolution to change our results significantly, 
however.  
First, it is the levelling off of the growth factor $D(t)$ at $z
\sim 1$ that results in the ISW effect.  Thus the bias is also relatively
constant during this epoch.
Second, the effect of bias evolution is the same as changing $dN/dz$, 
and our signal is not very sensitive to such changes.  
Scale dependent biasing, if arbitrary, makes any predictions for large scale
structure problematic.  However, assuming the scale dependence 
is relatively weak,  
then so long as we are measuring the radio bias on 
roughly the same scale as the expected correlations, our results should be 
fairly robust.   

\begin{figure}[htb]
\vspace{-0.4in} 
\centerline{
\psfig{file=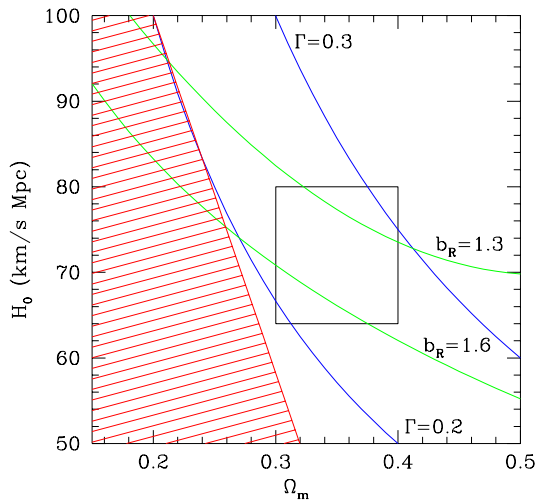,width=3.5in}}
\vspace{-0.3in} 
\caption{The region excluded at the 95\% CL 
in the Hubble constant-matter density plane. 
Also shown are contours of constant $\Gamma$.   The radio biases are 
inferred from the COBE normalization and the radio auto-correlation 
function.  The boxed region represents the observationally preferred region. 
}
\label{fig-omh}
\end{figure}

The integrated Sachs-Wolfe effect, if detected, would provide
an important confirmation of cosmological theory and 
provide a new mechanism to distinguish between models with a cosmological 
constant from quintessence dominated or open universes.  
Of course, if radio sources themselves contribute 
significantly to the CMB then such a correlation could be
explained otherwise.  However, current models
indicate that this is not the case.  From flux limited 1.4 GHz source
counts it can be shown that the contribution from these sources is less than
10\% of the expected ISW contribution to $\omega_{RT} (0)$   
and is even less for $\theta > 0$.
 
While the upper limit derived here is consistent
with the currently favored $\Omega_{\Lambda} \simeq 0.65$ CDM universe, it
is only marginally so.  
It is worth noting that a similar limit has been derived from the 
frequency of strong gravitational lensing \cite{gl}.  
Since these observations are so close to being in conflict with the 
leading cosmological model,  
it is important to
pursue both types of investigations further.  

The cross-correlation constraint will be greatly strengthened 
by NASA's recently launched Microwave Anisotropy Probe (MAP), which will 
provide a full sky map with much lower noise and higher
angular resolution than the COBE map.  Monte Carlos simulations
indicate that the cross-correlation of MAP with the NVSS survey
should be able to detect the ISW effect for an $\Omega_{\Lambda} = 0.6$
universe at the 95\% confidence level.  If, in addition, a future radio
survey is able to increase the number of sources by a factor of $\simge 3$,
then the Poisson counting noise will be effectively eliminated and
the detection should be at the $3 \sigma$ level.  Finally, if in the
future there becomes available a full-sky quasar or distant galaxy
redshift survey with several million objects, one will be able to
construct the `ideal' redshift distribution function and approach the
optimal $5.5 \sigma$ detection derived by Crittenden and Turok 
for a $\Omega_{\Lambda} = 0.6$ universe \cite{ct}.  
These will constitute crucial
tests which the $\Lambda$CDM universe must pass if it is to remain the
favored cosmological model.

We thank Neil Turok for useful discussions and Ed Groth for a variety
of analysis programs.  RC acknowledges support from a PPARC Advanced 
Fellowship.  This work was supported in part by NASA grant NAG5-9285.

\def\jnl#1#2#3#4#5#6{\hang{#1, {\it #4\/} {\bf #5}, #6 (#2).} }
\def\jnltwo#1#2#3#4#5#6#7#8{\hang{#1, {\it #4\/} {\bf #5}, #6; {\it
ibid} {\bf #7} #8 (#2).} } \def\prep#1#2#3#4{\hang{#1, #4.} }
\def\proc#1#2#3#4#5#6{{#1 [#2], in {\it #4\/}, #5, eds.\ (#6).} }
\def\book#1#2#3#4{\hang{#1, {\it #3\/} (#4, #2).} }
\def\jnlerr#1#2#3#4#5#6#7#8{\hang{#1 [#2], {\it #4\/} {\bf #5}, #6.
{Erratum:} {\it #4\/} {\bf #7}, #8.} } \def\prl{Phys.\ Rev.\ Lett.}
\def\pr{Phys.\ Rev.}  \def\pl{Phys.\ Lett.}  \def\np{Nucl.\ Phys.}
\def\prp{Phys.\ Rep.}  \def\rmp{Rev.\ Mod.\ Phys.}  \def\cmp{Comm.\
Math.\ Phys.}  \def\mpl{Mod.\ Phys.\ Lett.}  \def\apj{Astrophys.\ J.}
\def\apjl{Ap.\ J.\ Lett.}  \def\aap{Astron.\ Ap.}  \def\cqg{Class.\
Quant.\ Grav.}  \def\grg{Gen.\ Rel.\ Grav.}  \def\mn{Mon.\ Not.\ Roy.\
Astro.\ Soc.}
\def\ptp{Prog.\ Theor.\ Phys.}  \def\jetp{Sov.\ Phys.\ JETP}
\def\jetpl{JETP Lett.}  \def\jmp{J.\ Math.\ Phys.}  \def\zpc{Z.\
Phys.\ C} \def\cupress{Cambridge University Press} \def\pup{Princeton
University Press} \def\wss{World Scientific, Singapore}
\def\oup{Oxford University Press}

\end{document}